\documentclass[11pt,showpacs,preprintnumbers,superscriptaddress,amsmath,amssymb,nofootinbib]{revtex4}
\usepackage{graphicx}
\usepackage{dcolumn}
\usepackage{bm}
\usepackage{amssymb}
\usepackage{amsmath}
\usepackage{epsfig}    
\usepackage{color}
\usepackage{slashed}
\usepackage{hhline}

\def\be{\begin{equation}}
\def\ee{\end{equation}}
\def\Mat3#1#2#3#4#5#6#7#8#9{
\left(
\begin{array}{ccc}
#1 & #2 & #3 \\
#4 & #5 & #6 \\
#7 & #8 & #9 \\
\end{array}
\right) }
\newcommand{\bea}{\begin{eqnarray}}
\newcommand{\eea}{\end{eqnarray}}
\newcommand{\nn}{\nonumber}

\numberwithin{equation}{section}

\begin{document}

\title{Flavour Dependent Gauged Radiative Neutrino Mass Model}
\preprint{KIAS-P15001}
\author{Seungwon Baek}
\email{swbaek@kias.re.kr}
\affiliation{School of Physics, KIAS, Seoul 130-722, Korea}

\author{Hiroshi Okada}
\email{hokada@kias.re.kr}
\affiliation{School of Physics, KIAS, Seoul 130-722, Korea}

\author{Kei Yagyu}
\email{K.Yagyu@soton.ac.uk}
\affiliation{
School of Physics and Astronomy, University of Southampton, Southampton, SO17 1BJ, United Kingdom}

\begin{abstract}

We propose a one-loop induced radiative neutrino mass model 
with anomaly free flavour dependent gauge symmetry: $\mu$ minus $\tau$ symmetry $U(1)_{\mu-\tau}$. 
A neutrino mass matrix satisfying current experimental data 
can be obtained by introducing a weak isospin singlet scalar boson that breaks $U(1)_{\mu-\tau}$ symmetry, 
an inert doublet scalar field, and three right-handed neutrinos in addition to the fields in the standard model. 
We find that a characteristic structure appears in the neutrino mass matrix: two-zero texture form which 
predicts three non-zero neutrino masses and three non-zero CP-phases
from five well measured experimental inputs of two squared mass differences and three mixing angles. 
Furthermore, it is clarified that only the inverted mass hierarchy is allowed in our model. 
In a favored parameter set from the neutrino sector, 
the discrepancy in the muon anomalous magnetic moment between the
experimental data and the the standard model prediction can be 
explained by the additional neutral gauge boson loop contribution with mass of order 100 MeV and new gauge coupling of order $10^{-3}$.

\end{abstract}

\maketitle
\newpage

\section{Introduction}

Radiative neutrino mass models 
are one of the most promising scenarios at TeV scale physics 
to explain tiny neutrino masses. 
The original model based on the idea of radiative generation of neutrino masses is known as the Zee model~\cite{Zee} proposed in early 80's, where neutrino masses are 
generated at the one-loop level. 
After the Zee model, the Zee-Babu model~\cite{Zee-Babu} has also been proposed, where neutrino masses are explained at the two-loop level. 
In 2000's, radiative neutrino mass models have been extended so as to include a dark matter (DM) candidate by introducing an unbroken symmetry such 
as a discrete $Z_2$ symmetry known as; e.g., the models by Krauss-Nasri-Trodden~\cite{KNT} and by Ma~\cite{Ma,Ma_Pheno}. 
After these models appeared, various kinds of extensions have been considered in the scenario based on the raditive neutrino mass generation such as 
models with the supersymmetry~\cite{SUSY}, the B-L symmetry~\cite{B-L}, flavour symmetries~\cite{FlavorSym} and the conformal symmetry~\cite{Conformal}. 
Furthermore, in Refs.~\cite{Triplet}, a complex $SU(2)_L$ triplet scalar field is introduced, in which the collider phenomenology can be rich because of the 
existence of doubly-charged scalar bosons.   
Loop induced Dirac type neutrino masses have been proposed in Ref.~\cite{Dirac}. 
In models proposed in Refs.~\cite{radlepton}, charged lepton masses are also introduced at quantum levels in addition to neutrino masses. 
In addition to the above mentioned models, a number of radiative neutrino mass models have been constructed~\cite{Variation1,AKS,Variation2} up to now, and 
they have been classified into some groups in Refs.~\cite{Classify}.  

On the other hand, Abelian gauged $U(1)$ symmetries are well compatible with such radiative models. 
It has been known that 
there are four different anomaly free and flavour dependent types of $U(1)$ symmetries 
in the leptonic sector; namely, $L_{e}-L_{\mu}$, $L_{e}-L_{\tau}$,  and $L_{\mu}-L_{\tau}$, where $L_i$ denotes the lepton number with the flavour $i$. 
Especially in the case of  
$L_{\mu}-L_{\tau}$~\cite{Baek:2001kca, Chun:2007vh, baek, fox, Heeck:2011wj, He:1991qd, Foot:1994vd, Baek:2008nz, igari,Araki:2014ona}, 
constraints on the mass of additional neutral gauge boson $Z'$ and the new gauge coupling constant 
from the LEP experiment are very weak, because the $Z'$ boson does not
couple directly to the electron. 
We thus can consider a light $Z'$ boson scenario, by which 
the discrepancy in the muon anomalous magnetic
moment between current data and the prediction in the standard model (SM)~\cite{Baek:2001kca}
can be explained with the mass of $Z'$ to be ${\cal O}(100)$ MeV and the $U(1)_{\mu-\tau}$ gauge coupling to be ${\cal O}(10^{-3})$. 
The positron anomaly reported by AMS-02~ \cite{ams-02} could be explained~\cite{baek,Cao:2014cda}.
Such a light $Z'$ boson can be probed at the 14 TeV run of the LHC~\cite{igari}
through multi-lepton signals. 

In our paper, we combine a radiative neutrino mass model at 
one-loop level and the gauged $U(1)_{\mu-\tau}$ symmetry to get
neutrino masses, mixings, and dark matter candidates.  
We find that a predictive two-zero texture form of a neutrino mass matrix can be obtained corresponding to ``Pattern $C$'' in Ref.~\cite{twozero}. 
In this texture, we only need five experimental inputs to determine all the neutrino parameters. 
We can choose 
the most accurately measured ones: two squared mass differences and three mixing angles. 
It turns out that only the inverted mass hierarchy is allowed in our
texture.
Non-vanishing one Dirac and two Majorana CP-phases, and non-zero three
neutrino mass eigenvalues are predicted. 

This paper is organized as follows.
In Sec.~II, we define our model, and give mass formulae for scalar bosons.
In Sec.~III, we calculate the mass matrices for the lepton sector; charged leptons, right-handed neutrinos and left-handed neutrinos.
The detailed analysis for the two-zero texture form of neutrino mass matrix is also discussed. 
In Sec.~IV, we discuss new contributions to the muon $g-2$ and lepton flavour violation in our model. 
Conclusions and discussions are given in Sec.~V.

\section{The Model}

\begin{center}
\begin{table}[t]
\begin{tabular}{c||c|c|c||c|c|c}\hline\hline  
&\multicolumn{3}{c||}{Lepton Fields} & \multicolumn{3}{c}{Scalar Fields}  \\\hline
          &~$L_L^i=(\nu_L^i,e_L^i)^T$~ &~$e_R^i$~ &~$N_R^i$~ &~$\Phi$~  & ~$\eta$& ~$S$~   \\\hline 
$SU(2)_L$ & $\bm{2}$ & $\bm{1}$& $\bm{1}$ & $\bm{2}$&  $\bm{2}$ & $\bm{1}$ \\\hline 
$U(1)_Y$  & $-1/2$   & $-1$    & $0$     & $+1/2$    & $+1/2$   & $0$  \\\hline
$Z_2$ & $+$ & $+$ & $-$ & $+$& $-$& $+$ \\\hline\hline
\end{tabular}
\caption{The charge assignments of leptons and scalars under $SU(2)_L\times U(1)_Y$
 and $Z_2$ symmetry. The index $i (=e, \mu,\tau)$ denotes the lepton flavour. 
}
\label{tab1}
\vspace{1cm}
\begin{tabular}{c||c|c|c|c} \hline\hline
 & $(L_L^e,e_R,N_R^e)$ & $(L_L^\mu,\mu_R,N_R^\mu)$ & $(L_L^\tau,\tau_R,N_R^\tau)$ & $S$  \\ \hline
 $U(1)_{\mu-\tau}$ & $0$  &  $+1$  & $-1$ & $+1$    \\ \hline\hline
\end{tabular}
\caption{\label{tab2} The charge assignments under the gauged $U(1)_{\mu-\tau}$ symmetry. 
Fields which are not displayed in this table are neutral under $U(1)_{\mu-\tau}$. }
\end{table}
\end{center}


We consider a model in the framework of the gauge symmetry of
$SU(2)_L\times U(1)_Y\times U(1)_{\mu-\tau}$ with an unbroken discrete 
$Z_2$ symmetry. 
The particle content in our model is listed in Table~\ref{tab1}. 
The charge assignment for the $U(1)_{\mu-\tau}$ symmetry is separately shown in Table~\ref{tab2}. 

Our model is an extension of the model proposed by Ma~\cite{Ma}, where neutrino masses are generated at the one-loop level. 
In the Ma model,  
three right-handed neutrinos and an inert scalar doublet field are added to the standard model (SM). 
We introduce only one additional $SU(2)_L$ singlet scalar field $S$ with the even parity under $Z_2$ to the Ma model.
The vacuum expectation value (VEV) of $S$ breaks the $U(1)_{\mu-\tau}$ symmetry. 

The mass terms for right-handed neutrinos $N_R^i$ and the relevant Yukawa interactions are given by
\begin{align}
-\mathcal{L}_Y & = 
\frac{1}{2}M_{ee} \overline{N_R^{e\,c}} N_R^e + \frac{1}{2}M_{\mu\tau} (\overline{N_R^{\mu\, c}} N_R^\tau + \overline{N_R^{\tau\, c}} N_R^\mu)  +\text{h.c.} \notag\\
&+y_e \overline{L_L^e} \Phi e_R +y_\mu \overline{L_L^\mu} \Phi \mu_R + y_\tau \overline{L_L^\tau} \Phi \tau_R+ {\rm h.c.}\notag\\
&+h_{e\mu}^{}(\overline{N_R^{ec}}N_R^\mu+\overline{N_R^{\mu c}}N_R^e) S^* 
+ h_{e\tau}^{}(\overline{N_R^{e c}} N_R^\tau + \overline{N_R^{\tau c}} N_R^e )S + {\rm h.c.} \notag\\
&+f_e \overline{L_L^e}(i\sigma_2)\eta^* N_R^e  
 + f_\mu \overline{L_L^\mu} (i\sigma_2)\eta^* N_R^\mu 
 + f_\tau \overline{L_L^\tau} (i\sigma_2)\eta^* N_R^\tau +  {\rm h.c.} 
\label{yukawa}
\end{align}
The scalar sector of our model is composed of a singlet ($S$) and two
doublets, one active ($\Phi$) and one inert ($\eta$).  
The most general scalar potential is given by 
\begin{align}
\mathcal{V}
&= 
 \mu_\Phi^2 |\Phi|^2 + \mu_\eta^2 |\eta|^2  + \mu_S^2 |S|^2  \notag \\ 
&+\frac{1}{2}\lambda_1 |\Phi|^4 + \frac{1}{2}\lambda_2 |\eta|^4 + \lambda_3 |\Phi|^2|\eta|^2
+\lambda_4 |\Phi^\dagger \eta|^2
+\frac{1}{2}\lambda_5 [(\Phi^\dagger \eta)^{2} + \mathrm{h.c.}]\nn\\
&+\lambda_S^{} |S|^4 + \lambda_{S\Phi}^{} |S|^2 |\Phi|^2 + \lambda_{S\eta}^{}  |S|^2 |\eta|^2, 
\label{HP}
\end{align}
where all the parameters can be taken to be real without any loss of generality.
The scalar fields are parameterized by 
\begin{align}
\Phi = \left[\begin{array}{cc} 
G^+ \\ 
\frac{1}{\sqrt{2}}(v+\varphi_H^{}+iG^0)
\end{array}\right],\quad 
\eta = \left[\begin{array}{cc} 
\eta^+ \\ 
\frac{1}{\sqrt{2}}(\eta_H^{}+i\eta_A^{})
\end{array}\right],\quad 
S= \frac{1}{\sqrt{2}}(v_S^{}+S_H+iG_S),  \label{component}
\end{align}
where $v$ is the VEV related with the Fermi constant $G_F$ by
$v^2=1/(\sqrt{2}G_F)$, 
and $v_S^{}$ is the VEV of $S$ which breaks the $U(1)_{\mu-\tau}$ symmetry.  
In Eq.~(\ref{component}), 
$G^\pm$, $G^0$ and $G_S$ are the Nambu-Goldstone bosons which are absorbed by the longitudinal component of the $W^\pm$, $Z$ and 
an extra neutral gauge boson $Z'$ associated with the $U(1)_{\mu-\tau}$ symmetry, respectively. 

The tadpole conditions for $\varphi_H^{}$ and $S_H$ are respectively given by 
\begin{align}
&\frac{\partial \mathcal{V} }{\partial \varphi_H}\Big|_0 = v\left(\mu_\Phi^2 +\frac{v^2}{2}\lambda_1 + \frac{v_S^2}{2} \lambda_{S\Phi}\right) = 0, \notag\\
&\frac{\partial \mathcal{V} }{\partial S_H}\Big|_0 = v_S^{}\left(\mu_S^2 +\frac{v^2}{2}\lambda_{S\Phi} + v_S^2\lambda_{S}\right) = 0. 
\end{align}
Using the above two equations, we can eliminate $\mu_\Phi^2$ and $\mu_S^2$. 
There is no tadpole condition for $\eta_H^{}$, because the VEV of inert doublet field $\eta$ is zero due to the unbroken $Z_2$ symmetry. 

The $Z_2$-odd component scalar fields, $\eta^\pm$, $\eta_A$ and $\eta_H$, 
do not mix with the other fields, and their squared masses are simply given by  
\begin{align}
m_{\eta^\pm}^2 &= \mu_\eta^2 +\frac{v_S^2}{2}\lambda_{S\eta}+\frac{v^2}{2}\lambda_3, \\
m_{\eta_A}^2 &= \mu_\eta^2 +\frac{v_S^2}{2}\lambda_{S\eta}+\frac{v^2}{2}(\lambda_3+\lambda_4-\lambda_5), \\
m_{\eta_H}^2 &= \mu_\eta^2 +\frac{v_S^2}{2}\lambda_{S\eta}+\frac{v^2}{2}(\lambda_3+\lambda_4+\lambda_5). 
\end{align}
For the $Z_2$-even sector, two CP-even scalar states $\varphi_H^{}$ and $S_H$ are mixed with each other. 
Their mass matrix, $\mathcal{M}_H^2$, in the basis of $(\varphi_H^{},S_H)$ is given by 
\begin{align}
\mathcal{M}_H^2 = 
\begin{pmatrix}
v^2\lambda_1 & vv_S^{}\lambda_{S\Phi} \\ 
vv_S^{}\lambda_{S\Phi} & 2v_S^2\lambda_S 
\end{pmatrix}. \label{mass_mat}
\end{align}
The mass eigenstates for the CP-even states are given by introducing the mixing angle $\alpha$ by 
\begin{align}
\begin{pmatrix}
\varphi_H^{} \\ 
S_H
\end{pmatrix} = 
\begin{pmatrix}
\cos\alpha & -\sin\alpha \\ 
\sin\alpha & \cos\alpha
\end{pmatrix} 
\begin{pmatrix}
h \\ 
H
\end{pmatrix}. 
\end{align} 
In terms of the matrix element expressed in Eq.~(\ref{mass_mat}), the mass eigenvalues are 
\begin{align}
m_h^2 &= \cos^2\alpha (\mathcal{M}_H^2)_{11}+\sin^2\alpha (\mathcal{M}_H^2)_{22}+\sin2\alpha (\mathcal{M}_H^2)_{12}, \\
m_H^2 &= \sin^2\alpha (\mathcal{M}_H^2)_{11}+\cos^2\alpha (\mathcal{M}_H^2)_{22}-\sin2\alpha (\mathcal{M}_H^2)_{12}, 
\end{align}
and the mixing angle is 
\begin{align}
\tan2\alpha = \frac{2(\mathcal{M}_H^2)_{12}}{(\mathcal{M}_H^2)_{11}-(\mathcal{M}_H^2)_{22}}. 
\end{align}
We define $h$ as the SM-like Higgs boson with the mass of 126 GeV. 
Thus, $H$ corresponds to an additional singlet-like Higgs boson. 
Finally, if the conditions  
\begin{align}
&\lambda_1 > 0,\quad \lambda_2 > 0,\quad \lambda_S>0, \\
&\lambda_{S\Phi} + \frac{1}{\sqrt{2}}\sqrt{\lambda_1\lambda_S} > 0, \quad 
\lambda_{S\eta} + \frac{1}{\sqrt{2}}\sqrt{\lambda_2\lambda_S} > 0, \\
&\lambda_{3} + \frac{1}{2}\sqrt{\lambda_1\lambda_S}+\min(0,\lambda_4\pm\lambda_5) > 0. 
\end{align}
are satisfied, the Higgs potential Eq.(\ref{HP}) is bounded from below.

\section{Lepton Mass Matrix}

\begin{figure}[t]
\begin{center}
\includegraphics[scale=0.6]{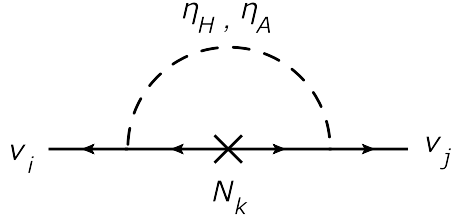}
\caption{Feynman diagram for neutrino masses at the one-loop level. In the internal fermion line, $N_k$ denotes the mass eigenstate of 
the right-handed neutrinos.  }
\end{center}
\end{figure}

The mass matrices for the charged-leptons and right-handed neutrinos are defined as 
\begin{align}
-\mathcal{L}_{\text{mass}}& = 
(\bar{e},\bar{\mu},\bar{\tau})\mathcal{M}_\ell(e,\mu,\tau)^T  \notag\\
&+ \frac{1}{2}(\overline{N_R^{e\,c}},\overline{N_R^{\mu\,c}},\overline{N_R^{\tau\,c}})\mathcal{M}_N (N_R^e,N_R^\mu,N_R^\tau)^T + \text{h.c.}, 
\end{align}
where $e,~\mu$ and $\tau$ are, respectively, $(e_L^{}+e_R^{}),~(\mu_L^{}+\mu_R^{})$ and $(\tau_L^{}+\tau_R^{})$. 
After the phase redefinition of the fields, $e_R^i$ and $N_R^i$, 
the mass matrices can be written in the form
\begin{align}
\mathcal{M}_\ell &= \frac{v}{\sqrt{2}}\text{diag}(|y_e|,|y_\mu|,|y_\tau|), \quad
\mathcal{M}_N  = 
\begin{pmatrix}
|M_{ee}| & \frac{v_S^{}}{\sqrt{2}}|h_{e\mu}| & \frac{v_S^{}}{\sqrt{2}}|h_{e\tau}| \\ 
\frac{v_S^{}}{\sqrt{2}}|h_{e\mu}|  & 0 & |M_{\mu\tau}|e^{i\theta_R}\\
\frac{v_S^{}}{\sqrt{2}}|h_{e\tau}| & |M_{\mu\tau}|e^{i\theta_R} & 0
\end{pmatrix},   \label{massmat}
\end{align}
where $\theta_R$ is the remaining unremovable phase.
Notice here that the $U(1)_{\mu-\tau}$ symmetry predicts the diagonal form of the mass matrix for the charged leptons. 
The mass matrix $\mathcal{M}_N$ is diagonalized by introducing a
unitary matrix $V$ satisfying
\begin{align}
 V^T \mathcal{M}_N V=\mathcal{M}_N^\text{diag} \equiv \text{diag}(M_{1},M_{2},M_{3}). 
\end{align}
The mass matrix for the left-handed Majorana neutrinos is then
calculated to be
\begin{align}
({\cal M}_\nu)_{ij}&=
\frac{1}{32\pi^2}\sum_{k=1\text{-}3}(f_iV_{ik}) M_{N_k} (f_jV_{jk})  
\left(\frac{m^2_{\eta_H}}{M^2_k-m^2_{\eta_H}}\ln\frac{m^2_{\eta_H}}{M^2_k}
-\frac{m^2_{\eta_A}}{M^2_k-m^2_{\eta_A}}\ln\frac{m^2_{\eta_A}}{M^2_k}
\right). 
\end{align}
If we assume $m^2_0\equiv (m^2_{\eta_H}+m^2_{\eta_A})/2 \gg M^2_k$,
the neutrino mass matrix can be simplified to be
\begin{align}
({\cal M}_\nu)_{ij}& \simeq 
-\frac{1}{32\pi^2} \frac{\lambda_5 v^2}{m^2_0}  \sum_{k=1\text{-}3}(f_iV_{ik}) M_{k} (f_jV_{jk}) \notag\\
&=-\frac{1}{32\pi^2} \frac{\lambda_5 v^2}{m^2_0} \sum_{k=1\text{-}3} f_i (\mathcal{M}_N)_{ij}f_j. 
\end{align}
More explicitly,  ${\cal M}_\nu$ can be written as 
\begin{align}
{\cal M}_\nu
&= \left( \begin{array}{ccc}
f_e^2M_{11}  & f_ef_\mu M_{12} & f_ef_\tau M_{13}  \\
f_ef_\mu M_{12}  & 0 & f_\mu f_\tau M_{23}e^{i\theta_R}  \\
f_ef_\tau M_{13} & f_\mu f_\tau M_{23}e^{i\theta_R}  & 0 
\end{array}\right), \label{mat2}
\end{align}
where we reparametrized dimension-full real parameters $M_{ij}$ defined as
\begin{align}
M_{11} = M_{ee},~M_{12} = \frac{v_S}{\sqrt{2}}h_{e\mu},~M_{13} = \frac{v_S}{\sqrt{2}}h_{e\tau},~M_{23} = M_{\mu\tau}, 
\end{align}
in the unite of $-\lambda_5v^2/(32\pi^2 m_0^2)$. 
The structure of matrix, Eq.~(\ref{mat2}), implies that the $U(1)_{\mu-\tau}$ symmetry predicts the so-called two-zero texture form of the Majorana neutrino mass matrix. 
Fifteen patterns of the two-zero texture form have been discussed in Ref.~\cite{twozero}, 
and our form corresponds to one termed ``Pattern $C$".
Because of the two zero texture form, nine neutrino parameters, three
mass eigenvalues, 
three mixing angles and three (one Dirac and two Majorana) CP-phases,
are predicted from five input parameters. 
In the following, we'll discuss how we can determine all the neutrino parameters by five experimental inputs. 

First, we introduce the Pontecorvo-Maki-Nakagawa-Sakata (PMNS) matrix
$U_{\text{PMNS}}$~\cite{pmns} 
to diagonalize the neutrino mass matrix: 
\begin{align}
\mathcal{M}_\nu & = U_{\text{PMNS}}\, {\rm diag}(m_1,m_2,m_3)\,U_{\text{PMNS}}^T , \label{mns0}
\end{align}
where $m_1$, $m_2$ and $m_3$ are the neutrino mass eigenvalues. 
The PMNS matrix is expressed as the product of two unitary matrices 
\begin{align}
U_{\text{PMNS}} = UP, 
\end{align}
where
\begin{align}
U \equiv \left[ \begin{array}{ccc}
1 & 0 & 0 \\
0 & c_{23}^{} & s_{23}^{}  \\
0 & -s_{23}^{} & c_{23}^{}
\end{array}\right]
\left[ \begin{array}{ccc}
c_{13}^{} & 0 & s_{13}^{}e^{-i\delta} \\
0 & 1 & 0  \\
-s_{13}^{}e^{-i\delta} & 0 & c_{13}^{} 
\end{array}\right]
\left[ \begin{array}{ccc}
c_{12}^{} & s_{12}^{} & 0 \\
-s_{12}^{} & c_{12}^{} & 0  \\
0 & 0 & 1
\end{array}\right], \quad P \equiv \text{diag}(e^{i\rho},e^{i\sigma},1),  \label{mns}
\end{align}
with $s_{ij}=\sin\theta_{ij}$ and $c_{ij}=\cos\theta_{ij}$. 
In Eq.~(\ref{mns}), $\delta$ is the Dirac phase, and $\rho$ and $\sigma$ are the Majorana phases. 
Using the matrix $U$, Eq.~(\ref{mns0}) is rewritten by 
\begin{align}
\mathcal{M}_{\nu}= U\, {\rm diag}(\tilde{m}_1,\tilde{m}_2,\tilde{m}_3)\,U^T , 
\end{align}
where $\tilde{m}_3 = m_3 e^{2i\rho}$, $\tilde{m}_2 = m_2 e^{2i\sigma}$ and $\tilde{m}_3 = m_3$. 

Second, we obtain the following two equations from the two-zero texture form 
\begin{align}
[U\, {\rm diag}(\tilde{m}_1,\tilde{m}_2,\tilde{m}_3)\,U^T]_{22} = [U\, {\rm diag}(\tilde{m}_1,\tilde{m}_2,\tilde{m}_3)\,U^T]_{33}= 0. 
\end{align}
This gives~\cite{twozero}
\begin{align}
\frac{\tilde{m}_1}{\tilde{m}_3} & = \frac{c_{12}^{}c_{13}^2}{s_{13}^{}}\frac{c_{12}^{}(c_{23}^2-s_{23}^2)e^{i\delta}-2s_{12}^{}s_{23}^{}s_{23}^{}c_{23}^{}}
{2s_{12}^{}c_{12}^{}s_{23}^{}c_{23}^{}(e^{2i\delta}+s_{13}^2)-s_{13}^{}(c_{12}^2-s_{12}^2)(c_{23}^2-s_{23}^2)e^{i\delta}}e^{2i\delta}, \notag\\
\frac{\tilde{m}_2}{\tilde{m}_3} & = -\frac{s_{12}^{}c_{13}^2}{s_{13}^{}}\frac{s_{12}^{}(c_{23}^2-s_{23}^2)e^{i\delta}-2s_{12}^{}s_{23}^{}s_{23}^{}c_{23}^{}}
{2s_{12}^{}c_{12}^{}s_{23}^{}c_{23}^{}(e^{2i\delta}+s_{13}^2)-s_{13}^{}(c_{12}^2-s_{12}^2)(c_{23}^2-s_{23}^2)e^{i\delta}}e^{2i\delta}. \label{ratio}
\end{align}
The ratios of neutrino mass eigenvalues and the Majorana phases are obtained from Eq.~(\ref{ratio}) as 
\begin{align}
& R_{13}\equiv \frac{m_1}{m_3} = \left|\frac{\tilde{m}_1}{\tilde{m}_3}\right|,\quad R_{23}\equiv \frac{m_2}{m_3} = \left|\frac{\tilde{m}_2}{\tilde{m}_3}\right|, \quad
\rho = \frac{1}{2}\text{arg}\left[\frac{\tilde{m}_1}{\tilde{m}_3}\right],\quad \sigma = \frac{1}{2}\text{arg} \left[\frac{\tilde{m}_2}{\tilde{m}_3}\right]. \label{ratio2}
\end{align}
Using $0\leq \theta_{ij}<\pi/2$ ($ij=12,~13$, and $23$) and $\theta_{13}\ll 1$,  we obtain  approximate formulae for $R_{13}$ and $R_{23}$ as 
\begin{align}
R_{13}& \simeq \left[1-\frac{2\cot\theta_{12}}{\sin\theta_{13}}\cot 2\theta_{23}\cos\delta +\left(\frac{\cot\theta_{12}}{\sin\theta_{13}}\cot2\theta_{23}\right)^2 \right]^{1/2}, \notag\\
R_{23}& \simeq \left[1+\frac{2\tan\theta_{12}}{\sin\theta_{13}}\cot 2\theta_{23}\cos\delta +\left(\frac{\tan\theta_{12}}{\sin\theta_{13}}\cot2\theta_{23}\right)^2 \right]^{1/2}.
\end{align}
In order to guarantee $m_2>m_1$ ( i.e., $R_{23}>R_{13}$), we require $\cot2\theta_{23}\cos\delta >0$.
In that case, we obtain $R_{23}>1$, which shows that only the inverted
mass hierarchy ($m_2>m_1>m_3$) is allowed 
in our model as already mentioned in Ref.~\cite{twozero}. 

Finally, 
we define the ratio of two squared mass difference: 
\begin{align}
R_\nu \equiv \frac{\Delta m_{21}^2}{|\Delta m_{31}^2|} = \frac{m_2^2-m_1^2}{|m_3^2-m_1^2|}. \label{rn1}
\end{align}
From Eq.~(\ref{ratio2}), it can be rewritten in the inverted mass hierarchy as
\begin{align}
R_\nu = \frac{R_{23}^2-R_{13}^2}{R_{13}^2-1}\simeq \frac{2}{\cos^2\theta_{12}}\frac{\cot2\theta_{12}\cot2\theta_{23}-\sin\theta_{13}\cos\delta}{2\sin\theta_{13}\cos\delta-\cot\theta_{12}\cot2\theta_{23}}. \label{rn2}
\end{align}
We can obtain three mass eigenvalues in terms of $\Delta m_{21}^2$,
$R_{13}$ and $R_{23}$ as
\begin{align}
m_3 = \frac{\sqrt{\Delta m_{21}^2}}{\sqrt{R_{23}^2-R_{13}^2}},\quad m_1 = m_3R_{31},\quad m_2 = m_3R_{23}. \label{rn3}
\end{align}

Now, we are ready to determine all the neutrino parameters by using five experimental inputs. 
The best fit (3$\sigma$ range) values in the inverted mass hierarchy are given as follows~\cite{nudata}: 
\begin{align}
&s^2_{12}= 0.323~(0.278\text{-}0.375),~~ s^2_{23} = 0.573~(0.403\text{-}0.640),~~ s^2_{13}=0.0229~(0.0193\text{-}0.0265), \notag\\
&\Delta m_{21}^2 = 7.60~(7.11\text{-}8.18)\times 10^{-5}~\text{eV}^2,~~ |\Delta m_{31}^2| = 2.38~(2.20\text{-}2.54)\times 10^{-3}~\text{eV}^2, 
\end{align}
In the following, we present our predictions using the three sets of input parameters; namely, 
using the best fit values (BF), using the upper limit of the 3$\sigma$ range ($+3\sigma$) and using the lower limit of the 3$\sigma$ range ($-3\sigma$). 
From two squared mass differences, we can obtain the numerical value
\begin{align}
 R_\nu = 0.0319~(\text{BF}),~ 0.0322~(+3\sigma),~ 0.0323~(-3\sigma), 
\label{Rnu}
\end{align}
from Eq.~(\ref{rn1}).
We can see that the analytic formula of $R_\nu$ in Eq.~(\ref{rn2}) is  a function of $\delta$. 
From Eq.~(\ref{rn2}) and Eq.~(\ref{Rnu}), we obtain the Dirac phase 
\begin{align}
\delta = \pm 1.96~(\text{BF}),~\pm2.07~(+3\sigma),~\pm 0.774~(-3\sigma).  
 \end{align}
All the negative (positive) solutions for $\delta$ are allowed (excluded) by the experimental data at 95\% CL~\cite{nudata}, so that we choose the negative solution. 
We then obtain the ratios as 
\begin{align}
&\frac{\tilde{m}_1}{\tilde{m}_3} = 1.39\times e^{1.91i},\quad
\frac{\tilde{m}_2}{\tilde{m}_3} = 1.40\times e^{-2.68i}~~(\text{BF}),\notag\\ 
&\frac{\tilde{m}_1}{\tilde{m}_3} = 2.03\times e^{1.52i},\quad
\frac{\tilde{m}_2}{\tilde{m}_3} = 2.06\times e^{-2.51i}~~(+3\sigma),\notag\\
&\frac{\tilde{m}_1}{\tilde{m}_3} = 1.73\times e^{-1.19i},\quad
\frac{\tilde{m}_2}{\tilde{m}_3} = 1.74\times e^{2.78i}~~(-3\sigma),
\end{align}
and the mass eigenvalues and Majorana phases from Eqs.~(\ref{ratio2}) and (\ref{rn3})
\begin{align}
&(m_1,m_2,m_3)~\text{[eV]} =  0.0583,~0.0589,~0.0420,~~(\rho,\sigma) = (0.956, -1.34)~~~(\text{BF}), \notag\\
&(m_1,m_2,m_3)~\text{[eV]} =  0.0533,~0.0540,~0.0262,~~(\rho,\sigma) = (0.759, -1.25)~~~(+3\sigma), \notag\\
&(m_1,m_2,m_3)~\text{[eV]} =  0.0585,~0.0591,~0.0339,~~(\rho,\sigma) = (-0.596, 1.39)~~~(-3\sigma).  
\end{align}
Using Eq.~(\ref{mns0}),  we can get the neutrino mass matrix $M_\nu$ 
\begin{align}
\mathcal{M}_\nu \simeq 
\begin{pmatrix}
0.0408 & -0.0186 & -0.0378 \\ 
-0.0186 & 0 & -0.0420-0.00631 i   \\
-0.0378 & -0.0420-0.00631 i & 0
\end{pmatrix}~\text{eV}~~~ \text{(BF)}, \notag\\
\mathcal{M}_\nu \simeq 
\begin{pmatrix}
0.0249 & -0.0228 & -0.0410 \\ 
-0.0228 & 0 & -0.0270-0.00352 i   \\
-0.0410 & -0.0270-0.00352 i & 0
\end{pmatrix}~\text{eV}~~~ (+3\sigma), \notag\\
\mathcal{M}_\nu \simeq 
\begin{pmatrix}
0.0321 & 0.0399 & 0.0271 \\ 
0.0399 & 0 & -0.0344+0.00252 i   \\
0.0271 & -0.0344+0.00252 i & 0
\end{pmatrix}~\text{eV}~~~ (-3\sigma), \label{mat3}
\end{align}
where we performed a phase redefinition so that the phase appears in the $(2,3)-$component as in Eq.~(\ref{mat2}).
We thus determine our model parameters by comparing each element of the above matrix with corresponding one given in Eq.~(\ref{mat2}).

We note in passing that the lightest right-handed neutrino can be a DM candidate in our scenario discussed in this section. 
The phenomenology of fermionic DM is quite similar to that in the Ma's model~\cite{Ma}, and its detailed discussions have been presented in Ref.~\cite{Ma_Pheno}. 

\section{Muon anomalous magnetic moment and lepton flavour violation}

The muon anomalous magnetic moment, so-called the muon $g-2$, has been 
measured at Brookhaven National Laboratory. 
The current average of the experimental results is given by~\cite{bennett}
\begin{align}
a^{\rm exp}_{\mu}=11 659 208.0(6.3)\times 10^{-10}. \label{obs}
\end{align}
It has been known that there is a discrepancy from the SM prediction by $3.2\sigma$~\cite{discrepancy1} 
to $4.1\sigma$~\cite{discrepancy2}:
\begin{align}
\Delta a_{\mu}=a^{\rm exp}_{\mu}-a^{\rm SM}_{\mu}=(29.0 \pm
9.0\ {\rm to}\ 33.5 \pm
8.2)\times 10^{-10}. \label{g-2_dev}
\end{align}
In our model, the dominant contribution to the muon $g-2$ is obtained through the one loop diagram where 
the muon and 
the extra neutral gauge boson $Z'$ of the $U(1)_{\mu-\tau}$ symmetry are running in the loop. 
The resulting form is given by
\begin{align}
\Delta a_\mu(Z^\prime) = \frac{g_{Z'}^2}{8\pi^2}\int_0^1 dx\frac{2 r x(1-x)^2}{r(1-x)^2+x}, 
\end{align}
where $g_{Z'}$ and $m_{Z'}$ are the $U(1)_{\mu-\tau}$ gauge coupling constant, the mass of $Z'$, respectively, 
and  $r\equiv (m_\mu/m_{Z'})^2$.

On the other hand, the parameter space on $m_{Z'}$ and $g_{Z'}$ has been severely constrained 
by the neutrino trident production process~\cite{neutrino_trident1} observed in 
neutrino beam experiments at the CHARMII~\cite{CHARMII} and at the CCFR~\cite{CCFR}, whose measured cross section
well agrees with the SM prediction.   
For example, 
$g_{Z'}\gtrsim 0.1$, $g_{Z'}\gtrsim 0.02$, $g_{Z'}\gtrsim 0.002$ and $g_{Z'}\gtrsim 0.001$ have been excluded with 95\% CL in the cases of 
$m_{Z'}=100$, 10 , 1 and 0.1 GeV, respectively~\cite{neutrino_trident1}. 
However, we note that the muon $(g-2)$ in our model is not constrained by the dark photon search experiment at BaBar because
$Z'$ does not couple to the electron in our case~\cite{Lees:2014xha,Baek:2014kna}.

By taking into account the constraint from the neutrino trident production, the discrepancy in 
the muon $g-2$ can be compensated to be less than 2$\sigma$ by $m_{Z'}\simeq 200$ MeV with $g_{Z'}\simeq 10^{-3}$~\footnote{In addition to the $Z'$ loop contribution, 
there is a negative contribution to the muon $g-2$ from the $\eta^\pm$ and $N_i$ $(i=$1-3) loop diagram. 
However, it can be neglected due to the assumption $M_k^2 \ll   m_{\eta^\pm}^2$ that provides two-zero texture in the neutrino sector. }

The lepton flavor violation also arises through the $\eta^\pm$ loop in our model.
The most stringent constraint is imposed by the MEG experiment: $\mathcal{B}(\mu\to e\gamma) < 5.7 \times 10^{-13}$ \cite{meg}.
The branching fraction is written by
\begin{align}
& {\cal B}(\mu\to e\gamma)\simeq (900\ {\rm GeV}^2)^2\times \left|  
\sum_{i=1\text{-}3} \frac{f_e f_\mu}{2m_{\eta^\pm}^2}  V_{1i} V^*_{2 i} G\left(\frac{M^2_i}{m^2_{\eta^\pm}}\right)  \right|^2. 
\end{align}
If we take $ \sum_{i=1\text{-}3}f_e f_\mu V_{1i}V^*_{2i}\lesssim {\cal O}(10^{-3})$ with $m_{\eta^\pm}^{}={\cal O}(1)$ TeV, we can avoid this constraint. 
Therefore, the anomaly in the muon $g-2$ can be well explained in the favored parameter region suggested from neutrino data and lepton flavour violation data. 

We note that our $Z'$ boson does not couple to the SM quarks, 
because it appears from the $U(1)_{\mu-\tau}$ gauge symmetry; i.e., 
it has a {\it quark phobic} nature. 
Therefore, any constraints for the $Z'$ boson using hadron events such as dijet searches cannot be applied. 

\section{Conclusions and Discussions}

We have constructed a one-loop induced radiative neutrino mass model in the gauge symmetry 
$SU(2)_L\times U(1)_Y\times U(1)_{\mu-\tau}$ with 
the unbroken discrete $Z_2$ symmetry.
In our model, three right-handed neutrinos are introduced in addition to the SM, 
and the scalar sector is composed of two isospin doublets, 
one inert and one active, and a $U(1)_{\mu-\tau}$ charged singlet.

We have shown that the $U(1)_{\mu-\tau}$ symmetry predicts a characteristic structure of the lepton mass matrices. 
First, the mass matrix of charged leptons is diagonal in the interaction basis. 
Second, the mass matrix of left-handed neutrinos is in the two-zero texture form if inert scalar bosons 
are much heavier than the right-handed neutrinos. 
The two-zero texture form of the neutrino mass matrix has been intensively studied in Ref. \cite{twozero}, and 
our model provides a texture with vanishing (2,2) and (3,3) elements,  
corresponding to ``Pattern $C$'' in \cite{twozero}. 
In this pattern, only the inverted mass hierarchy is allowed. 
And we only need five input experimental data to fix the neutrino mass matrix. We can choose 
the most accurately measured ones: two squared mass differences and three mixing angles. 
Using the best fit values of five observables, we obtained non-zero Dirac and Majorana CP-phases, 
and non-zero three neutrino mass eigenvalues.

We showed that the $Z'$-loop contribution to the muon $g-2$
can explain the discrepancy between the current experimental data and the SM prediction
if the $Z'$ mass is of ${\cal O}(100)$ MeV and the $U(1)_{\mu-\tau}$ gauge coupling of ${\cal O}(10^{-3})$, which has not been excluded
by the neutrino trident production process. 
The constraint from lepton flavour violation such as $\mu\to e\gamma$ can be avoided in the parameter space favored by 
the neutrino data and the muon $g-2$. 

Finally, we would like to briefly discuss the collider phenomenology of our model. 
One of the important features is that the SM-like Higgs boson $h$ which was discovered at the LHC~\cite{Higgs_ATLAS,Higgs_CMS} 
can mix with the $U(1)_{\mu-\tau}$ Higgs boson $H$ via the mixing angle $\alpha$. 
As a consequence, the coupling constants of $h$ with the SM gauge bosons $hVV$ ($V=W,Z$) and fermions $hf\bar{f}$ can be universally deviated from those of the SM predictions
by the factor $\cos\alpha$. 
Since the pattern of the deviation in the $h$ couplings strongly depends on the structure of the Higgs sector as discussed in Ref.~\cite{fingerprint}, 
we can indirectly probe the model by looking at the deviation even if we cannot discover new particles such as $H$. 
In future collider experiments such as the LHC Run-II, the high luminosity LHC and the International Linear Collider (ILC), 
the Higgs boson couplings are expected to be measured quite accurately, especially they can be measured. 
In particular at the ILC with the collision energy of 500 GeV and the integrated luminosity of 500 fb$^{-1}$, 
the $hVV$ and $hf\bar{f}$ ($f=b,\tau$ and $t$) couplings can be measured with about 0.4\% and $\mathcal{O}(1)\%$~\cite{ILC}, respectively. 
Therefore, we can test our model by the comparison between the precisely measured Higgs boson coupling and the theory predictions. 

\vspace{0.5cm}
\hspace{0.2cm} {\it Acknowledgments}

The authors would like to thank W. Altmannshofer, T. Araki, J. Heeck and M. Kohda
for a useful comment on the neutrino trident production related to the $Z'$ contribution to the muon $g-2$. 
This work was supported  in part by National Research Foundation of
Korea (NRF) Research Grant 2012R1A2A1A01006053 (SB).
KY was supported by JSPS postdoctoral fellowships for research abroad.


\end{document}